\documentstyle[12pt]{article}
\newcommand{\be}{\begin{equation}}
\newcommand{\ee}{\end{equation}}
\newcommand{\beas}{\begin{eqnarray*}}
\newcommand{\eeas}{\end{eqnarray*}}
\newcommand{\bea}{\begin{eqnarray}}
\newcommand{\eea}{\end{eqnarray}}
\newcommand{\bd}{\begin{description}}
\newcommand{\ed}{\end{description}}
\newcommand{\bi}{\begin{itemize}}
\newcommand{\ei}{\end{itemize}}
\newcommand{\bc}{\begin{center}}
\newcommand{\ec}{\end{center}}

\newcommand{\ra}{\rightarrow}
\newcommand{\lra}{\longrightarrow}

\newcommand{\half}{\frac{1}{2}}
\newcommand{\sfrac}[2]{\mbox{$\frac{#1}{#2}$}}
\begin{document}
\title{ Post-Newtonian Cosmology.}
\author{Peter Szekeres and Tamath Rainsford\\
Department of Physics and Mathematical Physics, \\
University of Adelaide, South Australia 5005, Australia}
\date{}
\maketitle

\begin{abstract}
 Newtonian Cosmology is commonly used in astrophysical
problems, because of its obvious simplicity when compared with general
relativity.  However it has inherent difficulties, the most obvious of
which is the non-existence of a well-posed initial value problem.  In this
paper we investigate how far these problems are met by using the
post-Newtonian approximation in cosmology.
\end{abstract}

\newpage

\section{Introduction}

In cosmology the use of Newtonian hydrodynamics is frequently favoured over
the more correct theory of general relativity \cite{bondi,heck-shuck,ellis}.
While the main reason for this is its obviously greater simplicity, there are
aspects of the Newtonian approximation which have distinct disadvantages.
In the first instance, it is not always understood what actually
constitutes the Newtonian approximation to general relativity.
Although generally assumed to be a specialization of the linear
approximation, this is not strictly true for
the Newtonian hydrodynamic equations 
are {\em not} the linearized approximation of the Bianchi identities,
since the term involving the gradient of the potential can only be obtained by
considering a higher (non-linear) order of approximation.

Secondly, and not totally unrelated to this difficulty, is the fact that the
equations of Newtonian hydrodynamics with gravitation do not have a well-posed
initial value problem.
Without a well-posed Cauchy problem the
future of initial fluctuations of density and velocity fields cannot be
uniquely determined from the Newtonian equations,
yet a major reason for using Newtonian theory in cosmology is that its
perturbation theory, both linearized and exact, is considerably easier than
that of general relativity \cite{bonnor,zeldo}.
What value are we to place then on the results of Newtonian perturbation theory
and its consequences for galaxy formation?

On the other hand it is well known that general relativity has a 
well-posed Cauchy problem for perfect fluids with an equation of state of the
form $P=f(\rho)$ \cite{synge,lich}.
What then goes wrong in taking the appropriate limit to Newtonian theory?
The principal aim of this paper is to 
write down a higher order approximation
of general relativity which results in a closed system, including Bianchi
identites, and has a well-posed Cauchy problem.
Although higher order approximations in general form an endless sequence
of equations \cite{wein,will}, called successively post-Newtonian, post-post
Newtonian etc., we will show that possible to make the cut-off at the 
post-Newtonian level in such a way that these goals are achieved. 

In Section 2 we outline the standard theory of Newtonian Cosmology and
describe in detail the difficulties stated above.
We then derive in Section 3 a standard sequence of approximations to general relativity.
It will be seen that wherever one curtails this sequence the Cauchy problem remains ill-posed.
In Section 4 the equations of the full post-Newtonian approximation are then 
cast in a form which gives a  closed system having a well-posed Cauchy problem.
In Section 5 we discuss future plans for this theory.

\section{Difficulties with Newtonian Cosmology}

Following the traditional approach of Heckmann and Sch\"{u}cking
\cite{heck-shuck,szek-rank}, a Newtonian cosmology is defined to be a
three dimensional Euclidean space and a universal time parameter $t$,
with a perfect fluid matter source for a scalar gravitational potential field.
Defined on the spacetime are three scalar functions, the density $\rho
(\vec{r},t)$, pressure $P(\vec{r},t) $ and
gravitational potential $\phi(\vec{r},t)$, and a velocity field
vector $\vec{v}(\vec{r},t) $.
These are related by the standard equations of fluid dynamics, namely
the equation of continuity, Euler's equation, and Poisson's equation:
\bea
\dot{\rho} +\vec{v} \cdot \nabla \rho + \rho \nabla \cdot\vec{v} = 0,
\label{Continuity}\\
\dot{\vec{v}} + \vec{v}\cdot \nabla \vec{v} = -\nabla \phi - \frac{1}{\rho}\nabla P,
\, \label{Euler}\\
\nabla^2 \phi =  4\pi G \rho.  \label{Poisson}
\eea
where throughout this paper a dot will refer to the partial derivative $\partial / \partial t$.

The system of equations (\ref{Continuity})-(\ref{Poisson}) together with an equation of state $P=f(\rho)$ is not well-posed however, as there is no proper initial value problem.  
Suppose one is given initial values $\rho_0(\vec{r})=\rho(\vec{r},t_0)$ and 
$\vec{v}_0(\vec{r})=\vec{v}(\vec{r},t_0)$, then clearly $P_0(\vec{r})$ is obtained from the equation of state, and the gravitational potential $\phi_0(\vec{r})$ can be found by solving equation (\ref{Poisson}).
Note however unlike standard Newtonian mechanics where the gravitational potential and its derivatives are always assumed to vanish asymptotically at large distances, in Newtonian {\em cosmology} no obvious boundary conditions at spatial infinity suggest themselves since the density is no longer assumed to vanish at infinity  
For this reason, the potential $\phi$ is only known up to an arbitrary solution of Laplace's equation $\nabla^2 \psi = 0$ (harmonic function).
While the time derivatives $\dot{\rho}_0$ and $\dot{\vec{v}}_0$ are determined from equations (\ref{Continuity}) and (\ref{Euler}), the time derivative of $\phi$ must also satisfy a Poisson equation,
\[\quad \nabla^2 \dot{\phi} =  4\pi G \dot{\rho}.\]
Again there arises an arbitrary harmonic function $\psi_1(\vec{r})$. 
This process continues for every higher time derivative of $\phi$, giving rise
to an infinite number boundary condition problems.  
The system of equations is not well-posed since $t = $ const is a characteristic surface of the system and it is not permitted to set initial data on such a surface \cite{courant-hilb}.

At first sight this is a peculiar phenomenon, since Newton's theory is a limiting approximation of general relativity, yet Einstein's equations are well known to have a well-posed Cauchy problem.
Some structural information is clearly being lost in taking this approximation.

The usual Newtonian theory is obtained from general relativity by
linearizing gravity and taking small characteristic velocities. 
In the weak field limit the metric has small deviations $\epsilon_{\mu \nu}$ from flat space and takes the form
\[g_{\mu \nu}=\eta _{\mu \nu} + \epsilon _{\mu \nu} \qquad \mbox{where} \quad
\epsilon=\left(\sum_{\mu}\sum_{\nu}\left|\epsilon_{\mu \nu}\epsilon_{\mu\nu}\right|\right)^{\frac{1}{2}} \ll 1.\]
As usual $\eta_{\mu \nu} =\mbox{diag} (-1,1,1,1)$, Greek indices run from 0 to 3 and Latin indices from 1 to 3. 

Since $g^{\mu \lambda}g_{\lambda \nu} = \delta ^{\mu}_{\nu}$, it follows that
\[g^{\mu \nu} = \eta^{\mu \nu} - \epsilon^{\mu \nu} + O(\epsilon^2), \qquad \mbox{where} \qquad \epsilon^{\mu \nu} = \eta^{\mu \rho}\eta^{\nu
\sigma}\epsilon_{\rho \sigma}.\]
Under an infinitesimal coordinate transformation
\[x^{\mu} = {x'}^\mu + \xi^{\mu}(x^{\alpha}),\]
with corresponding gauge transformation
\[\epsilon'_{\mu \nu} =\epsilon_{\mu \nu} + \xi_{\mu\,,\, \nu} + \xi_{\nu\,,\,\mu},\]
it is possible to choose coordinates $x^{\mu}$ for which the {\em harmonic gauge condition} holds,
\[\epsilon^{\alpha}{}_{\!\nu, \alpha} -\frac{1}{2}\epsilon^{\alpha}{}_{\!\alpha,\nu}= 0,\]
or equivalently,
\be \label{harmonic} \varphi^{\alpha}{}_{\!\nu, \alpha}= 0,\ee
where
\be \label{varphi}\varphi_{\mu \nu} = \epsilon_{\mu \nu} - \frac{1}{2} \eta_{\mu \nu} \epsilon^{\alpha}{}_{\!\alpha}.\ee 
In the harmonic gauge, the linearized Einstein equations read 
\be G_{\mu \nu} = -\frac{1}{2} {\varphi_{\mu \nu}{}^{,\alpha}}_{\!,\alpha} = \kappa T_{\mu\nu} \quad \mbox{where} \quad \kappa = 8\pi G c^{-4} \label{HEinstein} \ee

For a perfect fluid the energy-stress tensor takes the form
\be T_{\mu \nu} = (\rho c^2 +P)U_{\mu}U_{\nu} + Pg_{\mu \nu} \label{Pfluid}.
\ee
where, in the Newtonian approximation, the 4-velocity $U_{\mu}$ has components
\[U_{\mu} \approx (-1, \,\frac{v_i}{c}) + O(\beta^2),
\qquad\mbox{where}\quad \beta = v/c \ll 1.\]
The pressure $P$ will be assumed to be of order $O(\beta ^2)\rho c^2$ since
it is approximately the kinetic energy density, whence to highest order in each component
\[T_{00} =  \rho c^2 , \qquad T_{0i}=T_{i0}= -\rho v_i c , \qquad T_{ij}= \rho v_i v_j + P \delta_{ij} . \]  

Consistency in Einstein's equations (\ref{HEinstein}) can only be achieved in the limit $\beta \ra 0$ if it is assumed that $\epsilon$ and $\beta$ are the same order of magnitude, and
\[ \epsilon_{\mu\nu} \approx -2 \phi \delta_{\mu\nu} c^{-2} +O(\beta^4)\]
where $\phi$ is the Newtonian gravitational potential, assumed to be of magnitude $O(\beta^2 c^2)$.  
This is equivalent to assuming
\[ \varphi_{\mu\nu} = \mbox{diag}(-4\phi c^{-2}, 0,0,0) + O(\beta^4).\]
Finally, if for all physical quantities, derivatives $\partial/ \partial x^0$ are assumed to be of order $\beta \partial /\partial x^i$, then the $(0,0)$ component of Einstein's equations (\ref{HEinstein}) gives Poisson's equation (\ref{Poisson}).  

At this point we meet the first difficulty.
The remaining Newtonian equations (\ref{Continuity}) and (\ref{Euler}) should be obtained from the Bianchi identities.
However the linearized Bianchi identities are not consistent with these equations, since on using the harmonic gauge condition (\ref{harmonic}) it follows immediately that $T^{\mu \nu}{}_{\!, \nu} = G^{\mu \nu}{}_{\!,
\nu} = 0$. 
Substituting the above expression for the energy-stress tensor then results in an Euler equation {\em without} the term $\rho \phi_{,i}$.
The usual resolution of this problem is to include the term  $\Gamma ^i_{00} T^{00} \approx -\frac{1}{2} \rho c^2 \epsilon_{00,i}$ arising in the full (non-linear) Bianchi identites $ T^{\mu \nu}{}_{\!; \nu} = 0$. 

This mixture of linearized and non-linear approximations to general relativity is
clearly unsatisfactory.
It will be shown in Section 3 that by including the highest orders of approximation for the other 9 Einstein equations, which have so far been ignored, a version of Newtonian theory can be arrived at which yields the correct Bianchi identities.
While less familiar, this version of Newtonian theory contains precisely the same  information that is present in equations (\ref{Continuity})-(\ref{Poisson}).
It therefore resolves non of the difficulties discussed above.

We conclude this section with a brief discussion of the Cauchy problem in general relativity, and why it is that the Newtonian limit loses the well-posed character of Einstein's theory.
Following the treatment given by Synge \cite{synge} with minor modifications, a suitable initial data set given at $x^0 = 0$ is $g_{\mu \nu}, g_{ij,0}, \rho$ and $U_i$. 
The pressure is assumed given by an equation of state $P=f(\rho)$.
In order to obtain a unique solution of $G_{\mu \nu} = \kappa T_{\mu
\nu}$ it is necessary to subject this initial data to constraint equations
\be \label{constraint} G^0_{\,i}  =  \kappa T^0_{\,i}, \ee
and the evolution equations
\be \label{fullharmonic} g^{\rho\sigma}\Gamma^{\mu}_{\rho \sigma} = 0,\ee 
\be \label{fullbianchi}T^{\mu}_{\;\nu;\mu} = 0,\ee
\be \label{gijevol} R_{ij}  =  \kappa(T_{ij} - \frac{1}{2} T g_{ij}) \qquad \mbox{where} \quad T = T^{\alpha}_{\;\alpha}. \ee
Equation (\ref{fullharmonic}) is the full harmonic coordinate gauge condition and provides evolution equations for $g_{0\mu}$, while the Bianchi identities (\ref{fullbianchi}) are evolution equations for $\rho$ and $U_i$ which ensure that the constraint equations (\ref{constraint}) are maintained at later times.
The evolution of the metric components $g_{ij}$ are determined by Eqn (\ref{gijevol}) which has the structure
\[R_{ij} = -\mbox{$\half$} g^{00} g_{ij,00} + \mbox{ID,}\]
where ID refers to terms expressible entirely in terms of initial data quantities. 

When the Newtonian approximation is taken, $g_{\mu\nu} \approx \eta_{\mu\nu} - 2 \phi c^{-2} \delta_{\mu\nu}$, the only remaining equations at highest order 
$O(\beta^2)$ are as follows: the constraint equations reduce to the single Poisson's equation (\ref{Poisson}), the full Bianchi identities approximate to equations (\ref{Continuity}) and (\ref{Euler}) while the $R_{ij}$ evolution equation reduces again to Poisson's equation.  
Note how this latter evolution equation has become identical with the constraint equation in the Newtonian limit.
It is also worth observing that the {\em full} Bianchi identities are necessary to give the correct Euler equation.

\section{The Newtonian Approximation of a General Relativitistic Perfect Fluid}

Following a schema similar to that adopted by Weinberg \cite{wein}, we adopt units in which the typical velocity has magnitude 1, i.e.\ $\beta \approx c^{-1}$, and assume a one parameter family of metrics $g_{\mu\nu}(x^{\mu},c)$ for which there is a system of coordinates $(x^0, x^i)$ in which the components have the following asymptotic behaviour as $c \ra \infty$:
\bea
g_{00} & = & -1 -2 \phi c^{-2} - 2 \alpha c^{-4} - 2 \alpha' c^{-6} - 2
\alpha''c^{-8}..... \;, \nonumber \\
g_{0i} & = & \zeta_i c^{-3} + \zeta_i' c^{-5} + \zeta_i'' c^{-7}.....\;, \\
g_{ij} & = & \delta_{ij} - 2 \phi \delta_{ij} c^{-2} + \alpha_{ij} c^{-4} 
		+ \alpha_{ij}' c^{-6} + \alpha_{ij}'' c^{-8}.....\;.\nonumber
\eea

It will also be useful to expand the quantity $\varphi_{\mu \nu}$ defined in (\ref{varphi}), 
\beas 
\varphi_{00} & = &  -4 \phi c^{-2} + \theta c^{-4} + \theta' c^{-6} +
\theta'' c^{-8}..... \;,  \\
\varphi_{0i} & = & \zeta_i c^{-3} + \zeta_i' c^{-5} + \zeta_i'' c^{-7}.....\;, \\
\varphi_{ij} & = & \phi_{ij} c^{-4} + \phi_{ij}' c^{-6} + \phi_{ij}''c^{-8}.....\;,
\eeas
whence
\[\alpha =  - \frac{1}{4}(\theta + \phi_{kk}),\qquad \alpha'  =  - \frac{1}{4}(\theta' + \phi_{kk}'), \ldots, \]
\[\alpha_{ij}  =  \phi_{ij} + \frac{1}{2}\delta_{ij}(\theta - \phi_{kk}), \qquad \alpha_{ij}'  =  \phi_{ij}' + \frac{1}{2}\delta_{ij}(\theta' -
\phi_{kk}), \dots .\]

The Harmonic Gauge condition (\ref{harmonic}) gives a series of equations in successive powers of $c^{-2}$,
\bea
\dot{\phi} & = & - \frac{1}{4}\zeta_{i,i},\label{H1}\\
\dot{\zeta_i} & = & \phi_{ij,j}, \label{H2}\\
\dot{\theta} & = & \zeta_{i,i}', \label{H3}\\
\dot{\zeta_i'} & = & \phi_{ij,j}',\label{H4}\\
\ldots & \ldots & \ldots \nonumber
\eea

Replacing the quantities which define the perfect fluid by the expansions
\beas
\rho & \lra & \rho + \rho'c^{-2} + \rho''c^{-4} + ...\, ,\\
P & \lra & P + P'c^{-2} + P''c^{-4} + ...\, ,\\
U_{\mu}& \lra  &\left|U_0\right|(-1,\; v_ic^{-1} + v_i'c^{-3} + v_i''c^{-5} + ...),
\eeas
it follows from $U_{\mu}U^{\mu} = -1 $ that
\[U_0^{\:2} = 1 +(2\phi + v^2)c^{-2} + (2\alpha+6\phi v^2 + v^4 + 2\zeta_i v_i + 2v_i v'_i)c^{-4} + \cdots.\]
The energy-momentum tensor is defined (\ref{Pfluid}) as before, with an equation of state $P=f(\rho)$, which implies for higher pressure terms,
\be P'=f'(\rho)\rho',\quad P'' = f'(\rho) \rho'' +\mbox{$\half$} f''(\rho)(\rho'')^2, \quad \ldots \label{Pressure} \ee
where primes on the function $f$ refers to its derivatives.
Expanding the Ricci tensor as a power series in $c^{-1}$,
\[R_{\mu\nu} = {}^{(2)}\!R_{\mu\nu}c^{-2} + {}^{(4)}\!R_{\mu\nu}c^{-4}+ \cdots \]
and substituting in the Einstein field equations 
\[R_{\mu\nu}= 8\pi Gc^{-4}\left(T_{\mu\nu} - \sfrac{1}{2} T g_{\mu\nu} \right)\]
we find in the harmonic gauge, to order $c^{-4}$, 
\bea
{}^{(2)} \!R_{00} : \hspace{3cm} \phi_{,kk} & = & 4\pi G \rho, \label{R200}\\
{}^{(3)} \!R_{0i} : \hspace{3cm} \zeta_{i,kk} & = & 16\pi G\rho v_i, \label{R30i}\\
{}^{(4)} \!R_{ij} : \hspace{2.8cm} \phi_{ij,kk} & = &  - 16\pi G (\rho v_iv_j -\delta_{ij} P) + A_{ij} ,\label{R4ij} \\
{}^{(4)} \!R_{00} : \hspace{3.5cm}  \ddot{\phi} & = & \frac{1}{4}(-\theta_{,kk} +A) ,\label{R400}
\eea
where
\be A_{ij} \equiv 8\phi\phi_{,ij} + 4\phi_{,i}\phi_{,j} - \delta_{ij} (6\phi_{,k}\phi_{,k} + 32\pi G\rho\phi),\label{Aij} \ee
and
\be A \equiv 6\phi_{,i}\phi_{,i} - 16 \pi G (\rho v^2 + 4\rho\phi - \rho'). \label{A} \ee

The first three equations of this set (\ref{R200})--(\ref{R4ij}) together with 
the harmonic gauge conditions (\ref{H1}) and (\ref{H2}) constitute a 
reformulation of Newtonian cosmology since the time derivate of (\ref{R200}) 
and  $\partial / \partial x^i$ of (\ref{R30i}) together give rise to the equation of continuity (\ref{Continuity}), while $\partial / \partial t$ of (\ref{R30i})
and $\partial / \partial x^j$ of (\ref{R4ij}) give rise to Euler's equation (\ref{Euler}) with the correct gravitational term.

This version of Newtonian cosmology seems a little strange, since it introduces a new vector field $\zeta_i$ and tensor field $\phi_{ij}$.
It should be realized however that these two fields play an entirely {\em subsidiary} role in that the equations of Newtonian cosmology (\ref{Continuity}), 
(\ref{Euler}) and {\ref{Poisson}) are precisely the integrability conditions for (\ref{H1}), (\ref{H2}),  (\ref{R30i}) and (\ref{R4ij}).
The fields $\zeta_i$ and $\phi_{ij}$ play no further part in the theory.
In similar vein, although our version of Newtonian theory is a 4th order 
approximation to general relativity, the remaining 4th order equation 
(\ref{R400}) is also entirely subsidiary in nature, since it merely serves to 
define the next order approximation to density, $\rho'$, in terms of and arbitrarily specified quantity $\theta$.

Although equations (\ref{H1}) and (\ref{H2}) give rise to a time evolution equation for the gravitational potential,
\[ \ddot{\phi} = -\frac{1}{4} \phi_{ij,ij},\]
this in no way helps with the well-posedeness problem since in order to obtain 
higher time derivatives of $\phi$ it will be necessary to solve an infinite 
sequence of complicated Poisson-like equations for higher derivatives of 
$\phi_{ij}$ which arise on taking successive time derivatives of eq.(\ref{R4ij}).

\section{The Post-Newtonian Approximation}

Continuing the approximation of Einstein's equations to order $c^{-6}$, results in the equations
\bea
{}^{(5)} \!R_{0i}:\hspace{3cm} \ddot{\zeta}_i & = & \zeta'_{i,jj} + B_i, \label{R50j} \\
{}^{(6)} \!R_{ij} :\hspace{2.8cm} \ddot{\phi_{ij}}  & = &  \phi'_{ij,kk} + B_{ij}, \label{R6ij} \\
{}^{(6)} \!R_{00} : \hspace{2.85cm} \ddot{\theta}_{ij}  & = &  \theta'_{,kk} + C, \label{R600}
\eea
where
\bea B_i & \equiv &  3\zeta_{j,j}\phi_{,i} +
2\zeta_j\phi_{,ij} - 2\phi_{,j}\zeta_{j,i}\nonumber \\
& & \quad - 16\pi G\left[\rho v'_i + v_i(\rho' + P) + \rho v_iv^2 -
\sfrac{1}{2}\rho \zeta_i\right], \label{Bi} \\
B_{ij} & \equiv & -\sfrac{1}{2}\left(\zeta_i\zeta_{k,kj} + \zeta_j\zeta_{k,ki}\right) - \zeta_k\left(\zeta_{i,jk} + \zeta_{j,ik}\right) + 2\zeta_k\zeta_{k,ij} + \zeta_{k,i}\zeta_{k,j} + \zeta_{i,k}\zeta_{j,k}\nonumber \\ 
&& - 2\phi_{,k}\left(\phi_{ki,j} + \phi_{kj,i} - 2\phi_{ij,k}\right)
- 16\phi\phi_{,i}\phi_{,j} + 2\phi_{,i}\theta_{,j} + \phi_{,i}\phi_{mm,j} +
2\phi_{,j}\theta_{,i} \nonumber\\
& & + \phi_{,j}\phi_{mm,i} - 2\phi\left(\phi_{ki,jk}
+\phi_{kj,ik} - \phi_{ij,kk} - \phi_{mm,ij} - \theta_{,ij}\right) -
2\phi_{ki}\phi_{,jk} \nonumber\\
& &  - 2\phi_{kj}\phi_{,ik} + 2\phi_{,ij}(\theta + \phi_{mm})
- \delta_{ij}\left[\sfrac{1}{2}\zeta_{k,i}\zeta_{k,i} +
\sfrac{1}{2}\zeta_{j,k}\zeta_{k,j} + \sfrac{1}{2}(\zeta_{k,k})^2
\right.\nonumber\\
& &  - \zeta_k\zeta_{m,mk} - 4\phi_{,k}\phi_{ki,i} +
4\phi_{,k}\phi_{ii,k} - 12\phi\phi_{,k}\phi_{,k} + 4\phi_{,i}\theta_{,i} -
2\phi_{ki}\theta_{,ik}\Big)\nonumber\\ 
& & - \phi\left(2\phi_{ki,ik} - 2\phi_{ii,kk} -
\sfrac{3}{2}\theta_{,ii}\right) + 8\pi G\left[2(\rho' + P)v_iv_j + 2\rho
(v_iv_j' + v_jv_i')\right.\nonumber\\
& &  + 2\rho(2\phi + v^2)v_iv_j + \rho\phi_{ij}
+\delta_{ij}\left(2P' + 2\rho \phi v^2 - \sfrac{3}{2}\rho \theta -
\sfrac{1}{2}\phi \rho' - \sfrac{1}{2}\phi P\right.\nonumber\\
& & \left.\left. + \sfrac{3}{4}\phi_{,k}\rho_{,k} + \sfrac{1}{2}\rho
\phi_{kk}\right)\right], \label{Bij}
\eea 
and
\bea
C & \equiv & \sfrac{3}{2}\zeta_{k,i}\zeta_{k,i} + \sfrac{1}{8}(\zeta_{k,k})^2 -
\sfrac{1}{2}\zeta_{j,k}\zeta_{k,j} + 2\phi_{,k}\phi_{ii,k} -
20\phi\phi_{,k}\phi_{,k} + 4\phi_{,i}\phi_{,i} \nonumber\\
& & - 2\phi_{ki}\theta_{,ik} - \phi\left(2\phi_{ki,ik} + 2\phi_{ii,kk} +
\sfrac{5}{2}\theta_{,ii}\right) + 4\phi_{ki}\phi_{,ik}\nonumber\\
& & + 8\pi G \left[2\rho'' + \sfrac{3}{2}P\phi - \sfrac{3}{2}\phi\rho' + 
2\rho'v^2 + 2Pv^2- \sfrac{9}{4}\phi_{,k}\rho_{,k} \right.\nonumber\\
& & \left. + \rho\left(4v_i'v_i + 6v^2\phi
+ 2v^4 - \sfrac{1}{2}\phi_{ii} + \sfrac{3}{2}\theta - 8\phi ^2 +
4v_k\zeta_k\right)\right]. \label{C} 
\eea
Equations (\ref{R6ij}) and (\ref{R600}) require a certain amount of juggling before they can be cast in form given here.
While eq.\ (\ref{R6ij}) does provide an equation for $\ddot{\phi}_{ij}$, 
this inviting feature is counteracted by the fact that $\phi'_{ij}$ is not arbitrary but must satisfy the 4th order constraint equation
\[\nabla^2 \nabla^2 \phi'_{ij}= - \nabla^2 B_{ij} + \ddot{A}_{ij} -16 \pi G\left[(\rho v_i v_j)\ddot{} + \delta_{ij} \ddot{P}\right] \]
where all time derivatives on the right hand side can be reexpressed in terms 
of initial data quantities (undotted quantities, but possibly involving 
spatial derivatives) via the various evolution equations such as the harmonic 
conditions and Bianchi identities.
Without this constraint equations (\ref{R400}) and (\ref{R600}) will not be consistent with each other. 
Consequently all that has been achieved is to push the system to a higher order of accuracy, but with no further resolution of the initial value problem. 
Reading off successive approximants by ``peeling off'' the higher powers of
$c^{-1}$ in Einstein's equations continues the same problem to higher and
higher levels.
At no stage is it possible to close off the system in a self-consistent way.

There is however another method \footnote{We are indebted to D. Hartley for this proposal.}.
Instead of reading off each successive power as a separate equation, suppose we ``chop off'' Einstein's equations (\ref{HEinstein}) at successive levels 
$c^{-4}$, $c^{-6}$ etc., retaining the entire equation instead of peeling off
its component parts.
The order $c^{-4}$ theory is then the Newtonian approximation given in the 
previous section, while the order $c^{-6}$ theory replaces equations (\ref{R200})-(\ref{R6ij}) by the wavelike equations 

\begin{eqnarray}
\ddot\phi - c^2 \phi_{,kk} & = & -4\pi G \rho c^2 - \frac{1}{4}(\theta_{,kk}- A), \label{N1} \\
\ddot\zeta_i - c^2\zeta_{i,kk} & = & -16 \pi G\rho v_i c^2 + \zeta'_{j,kk} +
B_i ,  \label{N2i} \\
\ddot\phi_{ij} - c^2\phi_{ij,kk} & = &  \phi_{ij,kk}' +B_{ij} + c^2\left[16\pi G (\rho v_i v_j + \delta_{ij} P) - A_{ij} \right],
\label{N3ij}
\end{eqnarray}
where the expressions for $A$, $A_{ij}$, $B_i$ and $B_{ij}$ have all been defined above.

Using the harmonic condition (\ref{H1}) and (\ref{H2}) we obtain the following  Bianchi identities from (\ref{N1}), (\ref{N2i}) and (\ref{N3ij}):
\be \dot{\rho} + (\rho v_i)_{,i} + \frac{1}{16\pi G c^2} \left(\dot{\theta}_{,kk} - \zeta'_{j,jkk} - \dot{A} - B_{j,j} \right) = 0 \label{B1} \ee
\bea \lefteqn{\rho (\dot{v}_i + v_j v_{i,j} + \phi_{,i})+ P_{,i} + (\dot{\rho} + (\rho v_j)_{,j}) v_i  } \nonumber \\
& & = \frac{1}{16\pi G c^2} \left( \dot{\zeta}'_{i,jj} + \dot{B}_i - \phi'_{ij,jkk} - B_{ij,j} + c^2 A_{ij,j} \right) \label{B2i} \eea
These equations form a closed and well-posed system in the following sense: 
\begin{enumerate}
\item Set 10 {\em arbitrary functions} of space and time, $\phi'_{ij}(\vec{r},t)$, $\rho'(\vec{r},t)$ and $v'_i(\vec{r},t)$.
\item Rewrite eqs.\ (\ref{N1}) and (\ref{N2i}) as a set of 4 {\em constraint} equations
by substituting $\ddot{\phi}= -\frac{1}{4} \phi_{ij,ij}$ and $\ddot{\zeta}_i = \dot{\phi}_{ij,j}$ which follow from the gauge conditions (\ref{H1}) and(\ref{H2}),
\bea \phi_{,kk} & = & 4\pi G \rho + \frac{1}{4c^2}(\theta_{,kk}- \phi_{jk,jk} - A), \label{C1} \\
\zeta_{i,kk} & = & 16 \pi G\rho v_i  + \frac{1}{c^2} \left(-\zeta'_{j,kk} +
\dot{\phi}_{ij,j} - B_i \right).  \label{C2i} \eea
\item Finally there are a total of 18 {\em evolution equations}, consisting of (\ref{H1}), (\ref{H2}), (\ref{H3}), (\ref{H4}), (\ref{N3ij}), (\ref{B1}), and (\ref{B2i}).
\end{enumerate}

The system is well-posed, for given initial data $\phi_{ij}(\vec{r},0)$, $\dot{\phi}_{ij}(\vec{r},0)$, $\theta(\vec{r},0)$, $\zeta'_i(\vec{r},0)$, $\rho(\vec{r},0)$, $v_i(\vec{r},0)$ all arbitrary functions of $\vec{r}$,  and $\phi(\vec{r},0)$, $\zeta_i(\vec{r},0)$ subject to the constraint equations (\ref{C1}) and (\ref{C2i}), then the evolution equations determine a unique space-time dependence for  $\phi$, $\zeta_i$, $\phi_{ij}$, $\theta$, $\zeta'_i$, $\rho$ and $v_i$.  
What makes the system closed and self-consistent is the fact that (\ref{B1}) and the $,i$ derivative of (\ref{C2i}) imply $\partial/\partial t$ of the first constraint equation (\ref{C1}), while (\ref{B2i}) and $,i$ of eq.\ (\ref{N3ij}) imply the time derivative of (\ref{C2i}).  
Hence the constraint equations are carried forward in time as a consequence of the evolution equations and will automatically be true at later times if they hold at $t=0$.
It should be remarked finally that eq. (\ref{R600}) is redundant in that it can be regarded as merely acting to define $\rho''$  in terms of $\theta'$ and other quantities.
                                      
Although every choice of functions $\phi'_{ij}$, $\rho'$ and $v'_i$ leads to a viable post-Newtonian cosmological theory, it is obviously simplest to set them all to zero.
It also follows then from (\ref{Pressure}) that $P'=0$, while consistency with 
the harmonic conditions (\ref{H3}) and (\ref{H4}) is most simply maintained by setting $\theta=0$ and $\zeta'_i =0$. 
With these simplifications the Bianchi identities (\ref{B1}) and (\ref{B2i}) read
\bea
\lefteqn{\dot\rho \left(1 + \frac{v^2 - 4\phi}{c^2}\right) + (\rho v_j)_{,j}\left(1 + \frac{v^2}{c^2}\right) + \frac{1}{c^2} \left[
\rho\left(2v_j\dot v_j + 2v_jv_k v_{k,j} +\sfrac{1}{2}\zeta_{j,j}\right)\right.  }\nonumber\\
& & \left. - \frac{1}{2}\rho_{,j}\zeta_j + (v_jP)_{,j} +
\frac{1}{16 \pi G}(2\phi_{,i}\zeta_{i,jj} - 2\zeta_i\phi_{,jii} 
- 3\zeta_{i,i}\phi_{,jj}) \right] = 0,
\label{BB1}
\eea
and 
\bea
\lefteqn{\rho(\dot{v}_i + v_{i,j}v_j + \phi_{,i}) + P_{,i} =\frac{1}{16\pi G c^2} \left[ -(\dot{A} + B_{j,j})v_i + \right. }\nonumber \\
& &  \left. \dot{B}_i - B_{ij,j} - 2\phi(A_{,i} +\phi_{jk,ki}) - \phi_i(A + \phi_{jk.jk}) \right] \label{BB2i}
\eea

\section{Conclusion}

In this paper we have shown that standard Newtonian cosmological theory is inadequate; the Bianchi identities are not obtainable from the field equations 
and there is no well-posed initial value problem. 
By keeping higher order terms it was shown that Newtonian theory can be 
reformulated as a new theory where the Bianchi identities are consistent with the field equations.  
However this new Newtonian theory, although consistent, still does not have a well-posed initial value problem, and it is necessary to go to the post-Newtonian level in order to achieve a physically viable cosmological theory.

The usual Newtonian gravity leads to models which have significant dissimilarities to the corresponding general relativistic models \cite{sss}. 
In particular the theory of anisotropic homogeneous cosmologies is quite different in the two theories.  
In Newtonian cosmology these models can, in the case of dust $P=0$, all be classified by giving 5 arbitrary functions of time (the components of shear), while in general relativity it is necessary to classify models into the well-known nine Bianchi types.
It will be interesting to see if this new post-Newtonian theory can produce a theory of homogeneous models more in line with the general relativistic scheme.

\end{document}